\documentclass{IEEEoj}
\makeatletter
\let\labelindent\relax
\makeatother
\usepackage[inline]{enumitem}

\usepackage{amsmath,amssymb,amsfonts}
\usepackage{algorithmic}
\usepackage{graphicx,color}
\usepackage{textcomp}
\usepackage{booktabs}
\usepackage{multirow}
\usepackage{lmodern}
\usepackage{hyperref}

\def\BibTeX{{\rm B\kern-.05em{\sc i\kern-.025em b}\kern-.08em
    T\kern-.1667em\lower.7ex\hbox{E}\kern-.125emX}}
\def\OJlogo{\vspace{-10pt}\includegraphics[height=20pt]{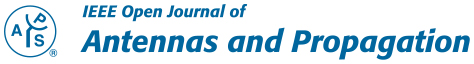}}

\newcommand{\dmin}{d_{\min}}
\newcommand{\Pth}{P_{\mathrm{th}}}
\newcommand{\Prx}{P_{\mathrm{rx}}}

\def\be{\begin{equation}}
\def\ee{\end{equation}}
\def\bea{\begin{eqnarray}}
\def\eea{\end{eqnarray}}

\newcommand{\orcid}[1]{\hspace*{2pt}
    \href{https://orcid.org/#1}{\includegraphics[width=10pt]{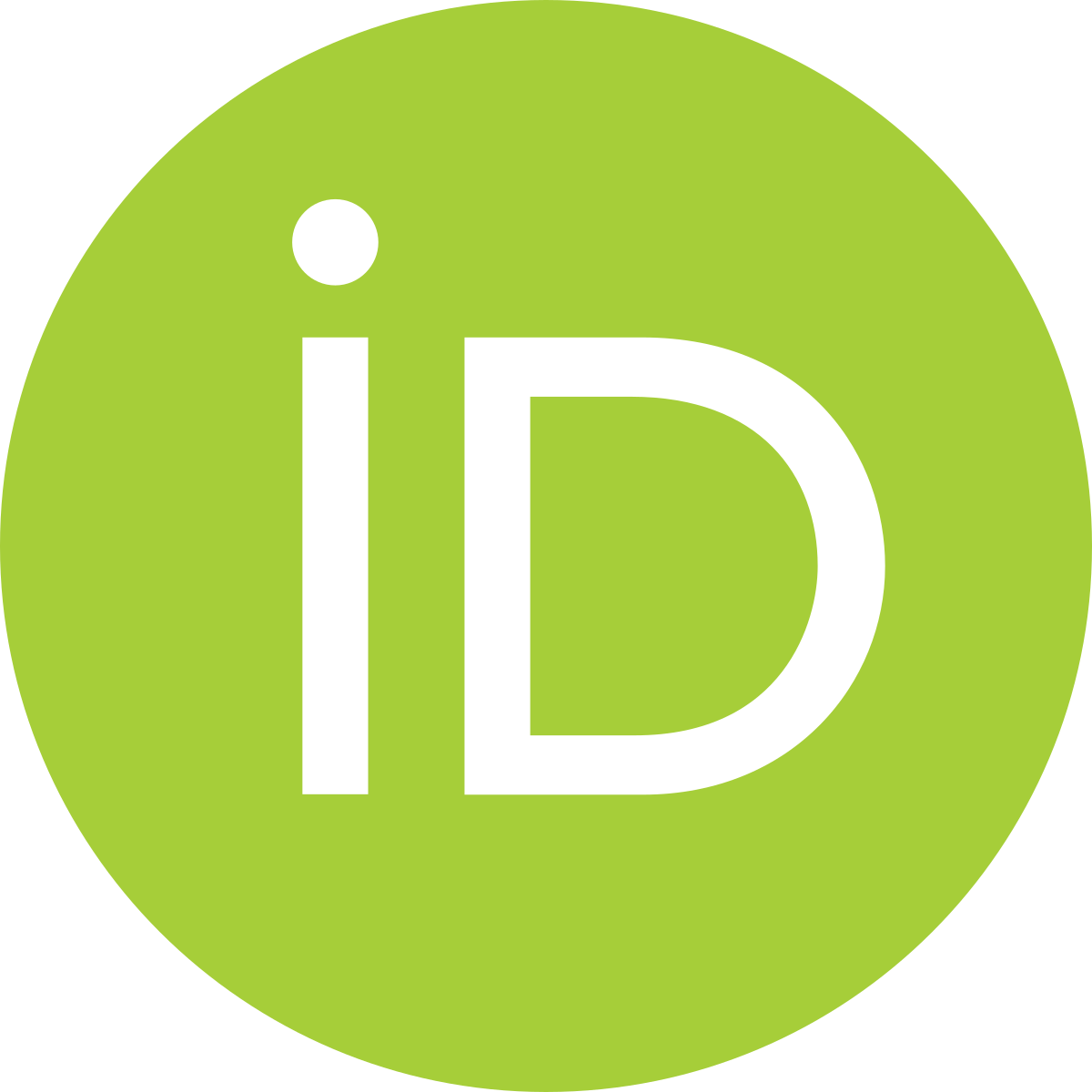}}}

\begin{document}


\title{A Comparative Systems-Engineering Framework for RFI Coexistence in Radio Astronomy and Aviation Safety Systems}

\author{SIMTHEMBILE DLAMINI\orcid{0000-0002-2885-6172}$^{\; 1\; 2 }$}
\affil{Department of Mathematical Sciences, University of South Africa, Florida, 1709, South Africa}
\affil{Centre for Astrophysics and Space Science, University of South Africa, Florida, 1709, South Africa}
\corresp{CORRESPONDING AUTHOR: Simthembile Dlamini (e-mail: simther4111@gmail.com)}
\authornote{This work received no external funding.}
\markboth{A Comparative Systems-Engineering Framework for RFI Coexistence}{Dlamini}

\begin{abstract}
Radio astronomy and aviation safety systems occupy opposite ends of the signal-power spectrum, yet both depend on the same finite resource, radio-frequency spectrum, and both are increasingly squeezed by the growth of commercial broadband wireless services. A central obstacle to treating this convergence rigorously is that the two fields protect their receivers using separately derived, apparently incommensurable criteria: the ITU-R Recommendation RA.769 detrimental-interference threshold for radio telescopes, and the radar-altimeter interference thresholds established by RTCA Special Committee 239 for the 5G C-band coexistence problem. We introduce a single coexistence-margin framework, $M(f,d)=\Pth(f)-\Prx(f,d)$, the protection threshold at a given frequency minus the interference power actually received at that frequency and separation distance, and show both criteria to be special cases of it. We apply this framework to two documented case studies, the legislated Karoo Radio Quiet Zone surrounding the MeerKAT and SKA-Mid telescopes in South Africa, and the global 5G C-band/radio-altimeter dispute in the 3.7--4.4 GHz region, demonstrating that, despite unrelated regulatory histories, both converged on the same three-lever solution: guard-band separation, bounded exclusion zones, and receiver-side filtering, governed by mandatory coordination rather than static exclusion. We then map the data-processing side of the problem: the interference-excision, beamforming, and statistical-calibration pipelines developed for radio interferometry are closely analogous to, and in several cases directly reusable for, the signal-processing operations required by ADS-B surveillance validation, phased-array radar clutter rejection, and aircraft predictive-maintenance anomaly detection. The unifying framework, and the finding that two independently-arising regulatory histories converge on the same design pattern under it, is this paper's central contribution: $M(f,d)$ has not, to our knowledge, been proposed before as a common formalism spanning these two protection regimes, and the technique-by-technique toolkit mapping in Section VI distinguishes genuine cross-domain reuse from looser analogy rather than asserting equivalence wholesale. The result is a reusable design pattern and a shared skills and infrastructure pipeline, applicable wherever a safety- or science-critical narrowband receiver must coexist with a growing broadband commercial neighbour.
\end{abstract}


\maketitle

\section{INTRODUCTION}
\label{sec:intro}

\IEEEPARstart{R}{adio} astronomy and civil aviation are not fields that usually appear
in the same technical literature, yet both operate receivers whose function depends on
detecting an extremely weak signal against a strong, growing, and largely uncoordinated
background of radio-frequency (RF) emissions. A radio telescope such as MeerKAT or the
Square Kilometre Array (SKA) must detect signals from the early universe that arrive at
power levels many orders of magnitude below the noise floor of a mobile
handset~\cite{Jonas:2016meerkat}. A radar altimeter on final approach must measure the
height of an aircraft above terrain to within centimetres, using a frequency-modulated
continuous-wave (FMCW) return signal that is comparably weak by the time it reaches the
receiver, and for which no redundant sensor exists during the final seconds before
touchdown~\cite{Bai:2025altimeter}. In both cases a missed detection means a lost
observation, or a lost aircraft.

\subsection{THE SHARED-VULNERABILITY PROBLEM}

This shared vulnerability is usually treated as a coincidence, worth a conference
anecdote, rather than a subject for formal comparison. That is the gap this paper
addresses. Both problems reduce to the same two engineering sub-problems, each with a
well-defined mathematical structure.

\begin{enumerate}[label=(\roman*), leftmargin=*]
  \item \textbf{The spectrum-protection problem.} Given a narrowband, safety- or
  science-critical receiver and a broadband commercial transmitter operating in an
  adjacent or shared band, define and enforce a quantitative condition under which the
  receiver is protected from harmful interference.
  \item \textbf{The big-data extraction problem.} Given a continuous, high-volume
  stream of noisy sensor data, in which the signal of interest is intermittently
  contaminated or entirely masked by interference, separate signal from
  interference/noise in (near-)real time and at a data rate that precludes exhaustive
  offline reprocessing.
\end{enumerate}

\subsection{A UNIFIED TREATMENT}

We formalise sub-problem (i) as a single coexistence margin, applicable without
modification to both a statistical protection criterion (radio astronomy) and a
functional, safety-of-flight threshold (radar altimetry); Section~\ref{sec:framework}
develops this formalism and shows the ITU-R RA.769 radio astronomy protection
criterion and the RTCA radar-altimeter interference-threshold methodology to be
special cases of it. Sections~\ref{sec:karoo} and~\ref{sec:cband} apply the formalism
to two documented case studies, the Karoo Radio Quiet Zone in South Africa and the
global 5G C-band/radio-altimeter dispute, and Section~\ref{sec:pattern} distils a
common \emph{coexistence design pattern} from the two. Section~\ref{sec:toolkit} then
turns to sub-problem (ii) and shows that the signal-processing toolkit built for radio
interferometry maps closely onto operational aviation data problems.
Section~\ref{sec:discussion} discusses the regulatory and skills-pipeline
implications, and Section~\ref{sec:conclusion} concludes.

\subsection{THIS PAPER}

This paper makes three contributions:
\begin{enumerate}[label=(\roman*), leftmargin=*]
  \item A single coexistence margin $M(f,d)=\Pth(f)-\Prx(f,d)$
  (Section~\ref{sec:framework}, Eq.~\ref{eq:margin}) that recovers the ITU-R RA.769
  criterion and the RTCA SC-239 threshold as domain-specific special cases.
  \item Two independently-arising regulatory histories
  (Sections~\ref{sec:karoo}--\ref{sec:cband}) that, despite no shared origin, converge
  on the same three-lever coexistence strategy, distilled into a general design
  pattern (Section~\ref{sec:pattern}).
  \item A transferable big-data signal-processing toolkit
  (Section~\ref{sec:toolkit}) relating RFI excision, beamforming, statistical
  calibration, and matched filtering/ML-assisted detection from radio interferometry
  to their closest aviation counterparts, distinguishing genuine technique reuse from
  looser conceptual analogy.
\end{enumerate}

\section{A UNIFIED ENGINEERING FRAMEWORK FOR NARROWBAND-RECEIVER PROTECTION}
\label{sec:framework}

\subsection{THE COMMON PROBLEM}

Stripped of domain-specific language, both radio astronomy and radar-altimeter
protection ask the same question: at what separation, in frequency and/or distance, does
an interfering transmitter stop being harmful to a specified receiver?\\~\\

We define a single \emph{coexistence margin}
\be
M(f,d) \;=\; \Pth(f) \;-\; \Prx(f,d) \qquad \text{[dB]},
\label{eq:margin}
\ee
where $\Pth(f)$ is the maximum interference power (or power spectral density, or power
flux density, depending on convention) that the protected receiver can tolerate at
frequency $f$ without a defined loss of function, and $\Prx(f,d)$ is the actual
interference power delivered to the receiver input by a transmitter located a
propagation distance $d$ away. Coexistence is achieved when $M(f,d) \geq 0$ with an
appropriate engineering safety margin; $M(f,d)<0$ denotes a harmful-interference
condition. The received power in a simple line-of-sight link budget is
\begin{equation}
\Prx(f,d) = P_{\mathrm{tx}} + G_{\mathrm{tx}} + G_{\mathrm{rx}}(\theta)
  - L_{\mathrm{fs}}(f,d) - L_{\mathrm{filt}}(f),
\label{eq:linkbudget}
\end{equation}
where all power and gain terms are in dBm (or dBW) and dB, respectively.\\~\\  
With free-space path loss
\begin{equation}
L_{\mathrm{fs}}(f,d) = 20\log_{10}(d) + 20\log_{10}(f) + 92.45,
\label{eq:fspl}
\end{equation}
where $d$ is in km and $f$ is in GHz,\\\\
$G_{\mathrm{rx}}(\theta)$ the receive-antenna gain in the direction of the
interferer (which for a radio telescope is normally the far side-lobe response, since
the main beam is rarely pointed at the interference source), and $L_{\mathrm{filt}}(f)$
any additional out-of-band rejection provided by front-end filtering. Equations
\eqref{eq:margin}--\eqref{eq:fspl} apply, without modification, to both domains; the two
fields differ only in how $\Pth(f)$ is defined and in which term of
equation~\eqref{eq:linkbudget} is engineered to restore $M\geq0$. Note that neither
domain treats $M(f,d)$ as a single global quantity: both evaluate it band-by-band, or
scenario-by-scenario, against a $\Pth(f)$ that is itself piecewise-defined.

\subsection{RADIO ASTRONOMY: THE ITU-R RA.769 CRITERION}
\label{sec:ra769}

For radio astronomy, $\Pth(f)$ is set by ITU-R Recommendation RA.769, the reference
standard used worldwide to justify spectrum protection for
observatories~\cite{ITU:RA769}. The Recommendation defines the detrimental
interference level as the power that would increase the RMS noise of a continuum
measurement, integrated over a standard $2000$\,s observation, by $10\%$, equivalent to
a $20\%$ loss of effective integration time~\cite{vanDriel:2009rfi}. For continuum
observations this criterion is tabulated, band by band, as a spectral power flux
density $S_H$; for example, in bands adjacent to the $2.7$\,GHz radio astronomy
allocation the threshold is $S_H=-247\,\mathrm{dB(W\,m^{-2}\,Hz^{-1})}$, corresponding
to an interference power of approximately $-177$\,dBm in a $10$\,MHz reference
bandwidth. Because radio telescopes typically observe with their main beam pointed away
from any interfering ground- or air-based transmitter, RA.769 conventionally assumes a
$0$\,dBi receive gain (i.e.\ the antenna's far side-lobe level) when evaluating
$G_{\mathrm{rx}}(\theta)$ in equation~\eqref{eq:linkbudget}~\cite{ITU:RA769}. This is
the RA.769 realisation of $\Pth(f)$ in equation~\eqref{eq:margin}: it is a
\emph{statistical} threshold, defined in terms of tolerable loss of sensitivity over an
integration, rather than a hard failure condition.

\subsection{AVIATION: RADAR-ALTIMETER INTERFERENCE THRESHOLDS}
\label{sec:altimeter}

For radar altimeters, $\Pth(f)$ is instead a \emph{functional} threshold: the injected
interference power at which a frequency-modulated continuous-wave (FMCW) altimeter's
height estimate is measurably degraded or lost. RTCA Special Committee 239 (SC-239)
established this experimentally for the 5G C-band coexistence problem, characterising
interference thresholds for fielded altimeters and finding that fundamental 5G
emissions in the $3.7$--$3.98$\,GHz band exceed safe interference limits for every
aircraft usage category under essentially all tested base-station and geometry
combinations~\cite{RTCA:SC239}. Subsequent laboratory work has quantified altimeter
interference thresholds directly: one recent study measured FMCW altimeter interference
thresholds of approximately $1$ to $6$\,dBm and $-4$ to $0$\,dBm at 5G centre
frequencies of $3.7$ and $3.9$\,GHz respectively~\cite{Bai:2025altimeter}. Companion
work from the same group has modelled the nonlinear intermediate-frequency gain
compression through which adjacent-band 5G emissions actually degrade the altimeter
receiver~\cite{Duan:2025eie}, and has proposed deployment-side mitigation, defining
horizontal and vertical exclusion zones around base stations and combining transmit
power control with antenna angle restrictions to keep $M(f,d)\geq0$ at the aircraft
without a blanket ban on nearby 5G transmission~\cite{Duan:2024deployment}. Unlike the
RA.769 criterion, this is a near-instantaneous, safety-of-flight threshold: there is no
equivalent of ``losing $20\%$ of integration time'' for an aircraft on short final,
which is why the aviation case has historically been resolved through hard exclusion
and retrofit mandates rather than statistical coordination.

\subsection{COMPARING THE TWO MARGINS}

Table~\ref{tab:margin_compare} summarises how the general framework of
equations~\eqref{eq:margin}--\eqref{eq:fspl} specialises in each domain. The structural
identity of the two problems is precisely why the same three engineering levers, namely
frequency separation ($f$ dependence of $L_{\mathrm{filt}}$), spatial separation ($d$
dependence of $L_{\mathrm{fs}}$), and receiver hardening ($G_{\mathrm{rx}}$ or
$L_{\mathrm{filt}}$), recur independently in both case studies examined in
Sections~\ref{sec:karoo} and~\ref{sec:cband}.

\begin{table*}[!t]
\centering
\caption{The coexistence-margin framework specialised to radio astronomy and to radar-altimeter protection. Both are instances of \texorpdfstring{$M(f,d)=\Pth(f)-\Prx(f,d)$}{M(f,d) = Pth(f) - Prx(f,d)}.}
\label{tab:margin_compare}
\begin{tabular}{p{3.1cm} p{5.7cm} p{5.7cm}}
\toprule
 & \textbf{Radio astronomy} & \textbf{Radar altimeter (5G C-band)} \\
\midrule
Governing standard & ITU-R RA.769~\cite{ITU:RA769} & RTCA SC-239 / national
airworthiness directives~\cite{RTCA:SC239,FAA:AD2021} \\
Nature of $\Pth$ & Statistical: $10\%$ RMS-noise increase over a $2000$\,s
integration & Functional: threshold at which height-estimate integrity is lost \\
Typical $\Pth$ & $S_H\!\approx\!-247\,\mathrm{dB(W\,m^{-2}\,Hz^{-1})}$ (example band)
& $\approx -4$ to $+6$\,dBm at receiver input~\cite{Bai:2025altimeter} \\
Assumed $G_{\mathrm{rx}}$ & $0$\,dBi (far side-lobe reference)~\cite{ITU:RA769} &
Front-end bandpass roll-off, $\sim\!24$\,dB/octave below $4.2$\,GHz~\cite{MiniCircuits:filter} \\
Primary mitigation levers & Legislated exclusion zones, coordination
procedures & Guard band ($220$\,MHz), exclusion zones near runways, filter retrofits \\
Consequence of $M<0$ & Increased integration time / data loss & Loss of a
safety-critical sensor during a critical flight phase \\
\bottomrule
\end{tabular}
\end{table*}

\begin{figure*}[!t]
\centering
\includegraphics[width=0.85\textwidth]{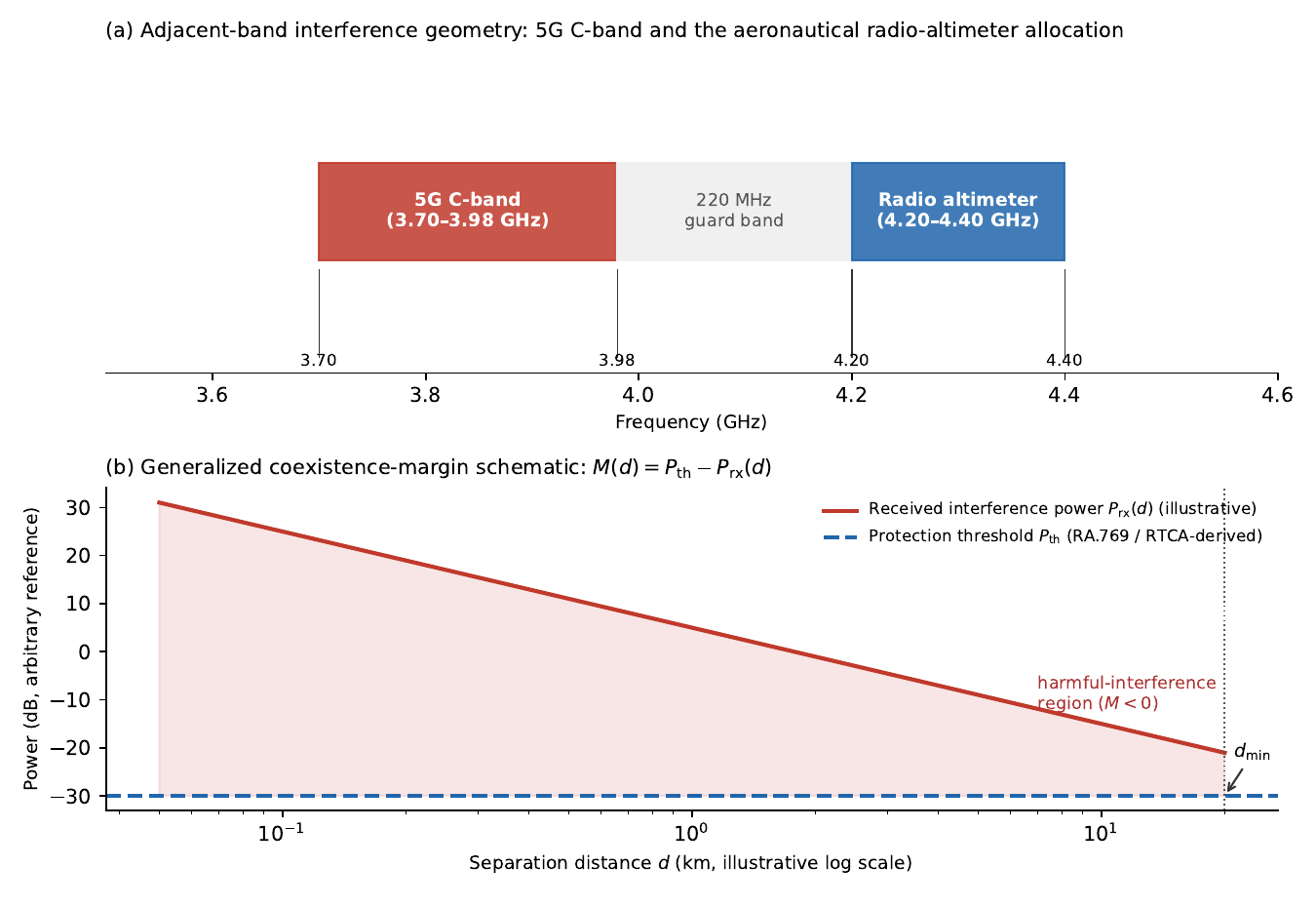}
\caption{(a) The real adjacent-band geometry underlying the 5G C-band / radar-altimeter
interference problem: the C-band mobile allocation ($3.70$--$3.98$\,GHz), the $220$\,MHz
guard band, and the aeronautical radio-altimeter allocation ($4.20$--$4.40$\,GHz)
identified by RTCA SC-239~\cite{RTCA:SC239}. (b) A schematic (illustrative, not
quantitatively fitted) rendering of the general coexistence margin
$M(d)=\Pth-\Prx(d)$ from equation~\eqref{eq:margin}: interference power falls off with
separation distance $d$ while the protection threshold is fixed, defining a minimum
coordination distance $\dmin$ below which the interference condition is harmful.}
\label{fig:coexistence}
\end{figure*}

Fig.~\ref{fig:coexistence} illustrates the two components of this framework
together: panel (a) shows the real frequency geometry of the 5G C-band /
radio-altimeter problem, and panel (b) shows, schematically, how the margin
$M(d)=\Pth-\Prx(d)$ changes with separation distance and defines a minimum coordination
distance $\dmin$, the general quantity that, in the radio astronomy case, is fixed by
legislation (Section~\ref{sec:karoo}), and in the aviation case is fixed by
airport-specific exclusion zones and NOTAM-defined restrictions
(Section~\ref{sec:cband}).

\section{CASE STUDY I: THE KAROO RADIO QUIET ZONE}
\label{sec:karoo}

South Africa's Astronomy Geographic Advantage Act 21 of 2007 established the Karoo
Central Astronomy Advantage Area, a protected zone in the Northern Cape surrounding
the MeerKAT array and the SKA-Mid telescope, now recognised as the largest radio quiet
zone on Earth at roughly $106{,}000\,\mathrm{km^2}$~\cite{AGA:2007}. In the language of
Section~\ref{sec:framework}, the Act does not attempt to drive $\Prx\to-\infty$
everywhere within the zone (which would require a total transmission ban); instead it
raises $L_{\mathrm{fs}}(f,d)$ and restricts $P_{\mathrm{tx}}$ for licensees within
successive coordination radii, so that $M(f,d)\geq0$ is satisfied at the telescope for
the RA.769 threshold appropriate to each protected band. South African Radio Astronomy
Observatory (SARAO) representatives have been explicit that the intent is mitigation
rather than elimination of interference, and the same coordination requirement now
extends to South Africa's broader satellite spectrum reforms, which oblige operators to
coordinate with the relevant authority to avoid interference with the Karoo
installations~\cite{ICASA:spectrum,SKAO:profile,SARAO:rqz}.

Two features of this case are directly relevant to the general framework:

\begin{enumerate}[label=(\roman*), leftmargin=*]
  \item \textbf{Zoned, not binary, protection.} The Act defines nested regions with
  different restriction levels rather than a single hard boundary, which is the
  legislative analogue of designing separate margins $M(f,d)$ for different classes of
  transmitter (broadcast, cellular, licensed point-to-point) rather than a single
  blanket rule.
  \item \textbf{Coexistence with safety-critical aviation, not exclusion of it.} Flight
  paths, radar operations, and airport communications within the region are
  coordinated, not excluded, so that aviation's own safety-critical spectrum use and
  radio astronomy's protection requirement are jointly satisfied. This is the clearest
  existing example of the general framework being applied \emph{symmetrically}: both
  parties hold a protected receiver, the telescope and the aircraft's own systems
  respectively, and coordination resolves the conflict.
\end{enumerate}

\begin{figure}[!t]
\centering
\includegraphics[width=0.85\columnwidth]{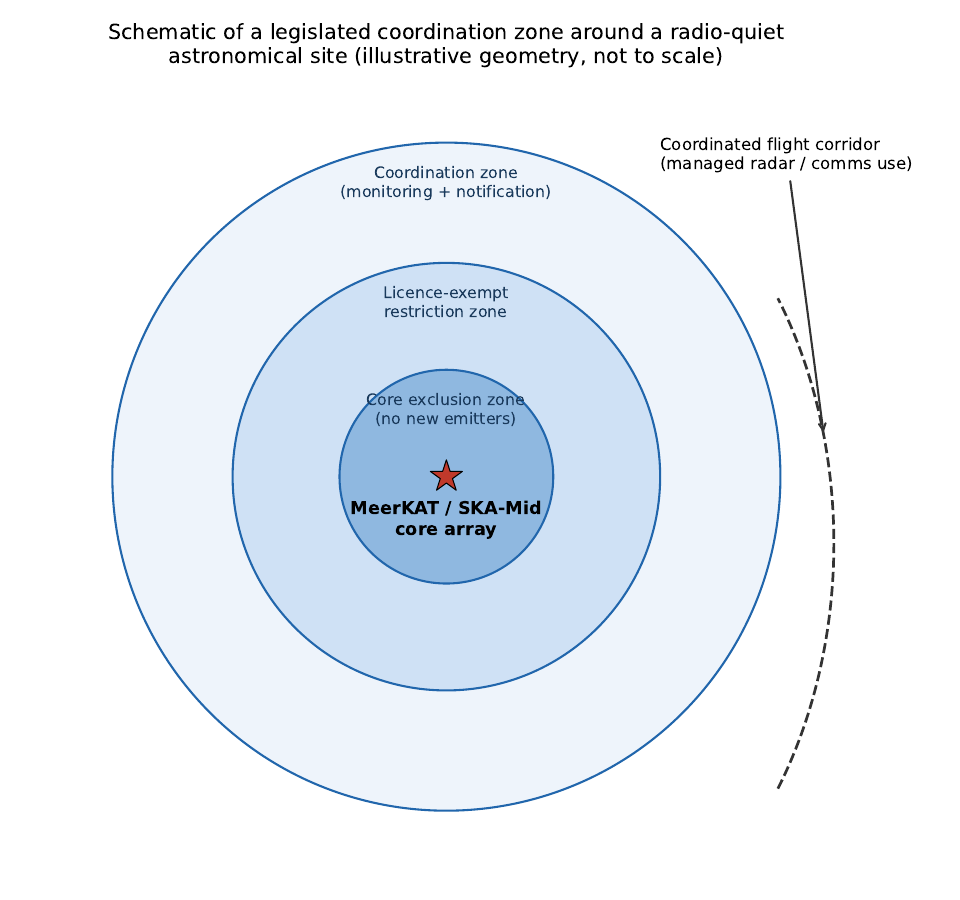}
\caption{Schematic representation of a legislated, tiered coordination zone around a
radio-quiet astronomical site, of the type established by the Astronomy Geographic
Advantage Act around the Karoo Central Astronomy Advantage Area~\cite{AGA:2007}. The
zone geometry shown is illustrative and not to scale; it is intended to convey the
structural point that protection is graded and coordinated rather than a single
exclusion boundary, and that safety-critical aviation traffic is accommodated through
coordination rather than exclusion.}
\label{fig:rqz}
\end{figure}

Fig.~\ref{fig:rqz} shows this structure schematically as a set of nested coordination
radii around the telescope core, together with an illustrative coordinated flight
corridor, mirroring the tiered radio-quiet-zone regulations used in comparable
observatory protection regimes internationally (e.g.\ the roughly $13{,}000$\,square-mile
US National Radio Quiet Zone surrounding the Green Bank Observatory and the Sugar
Grove naval facility~\cite{NRAO:nrqz}).

\section{CASE STUDY II: THE 5G C-BAND / RADIO-ALTIMETER PROBLEM}
\label{sec:cband}

The mirror-image case is aviation's own spectrum conflict. Radio altimeters operate at
$4.2$--$4.4$\,GHz, immediately adjacent to the $3.7$--$3.98$\,GHz C-band spectrum sought
globally for 5G deployment because of its favourable coverage-versus-capacity
trade-off~\cite{RTCA:SC239}. RTCA SC-239's assessment found that, for commercial and
general-aviation aircraft categories, harmful interference from 5G base stations was
essentially unavoidable across the tested range of transmitter configurations and
aircraft-to-base-station geometries, driven by two mechanisms: fundamental emissions in
the C-band itself, and spurious/out-of-band emissions leaking into the altimeter
band~\cite{RTCA:SC239}. This is directly attributable, in the language of
equation~\eqref{eq:linkbudget}, to altimeter front-end filters that were only ever
specified to roll off at $\sim\!24$\,dB per octave beyond their design band, because the
C-band had for decades been occupied only by low-power satellite links. That filtering
budget was adequate for the interference environment altimeters were designed against,
but inadequate once $P_{\mathrm{tx}}$ in the adjacent band rose by many orders of
magnitude with terrestrial 5G deployment~\cite{MiniCircuits:filter}.

The resolution, as in the Karoo case, combined the same three levers identified in
Section~\ref{sec:framework}:

\begin{itemize}
  \item \textbf{Frequency-domain separation:} a $220$\,MHz guard band between the C-band
  edge ($3.98$\,GHz) and the altimeter band edge ($4.2$\,GHz), together with power
  limits on base stations operating near that edge;
  \item \textbf{Spatial/temporal exclusion:} geographically bounded restriction zones
  around major runways, and operational limitations published via Notices to Air
  Missions (NOTAMs) and incorporated into aircraft flight manuals by airworthiness
  directive~\cite{FAA:AD2021};
  \item \textbf{Receiver hardening:} an industry-wide retrofit programme fitting
  altimeters with sharper, higher-rejection bandpass filters to restore adequate
  $L_{\mathrm{filt}}(f)$ against the new interference environment~\cite{MiniCircuits:filter}.
\end{itemize}

By the mid-2020s, aviation authorities and telecom regulators in major markets had
reached a stable coexistence framework combining these three levers, though the process
took years, required retrofit costs across entire fleets, and remains under active
refinement in some jurisdictions as further C-band spectrum is allocated.

\section{A GENERALIZED COEXISTENCE DESIGN PATTERN}
\label{sec:pattern}

The two case studies, despite arising from unrelated regulatory histories and unrelated
technical communities, converge on the same design pattern, which we state generally:

\begin{quote}
\textbf{Coexistence design pattern.} Given a protected receiver with threshold
$\Pth(f)$ and an adjacent or co-band transmitter with power $P_{\mathrm{tx}}$, achieve
$M(f,d)\geq0$ (equation~\ref{eq:margin}) through a combination of (1) frequency-domain
separation (guard bands, power spectral masks), (2) spatial and/or temporal exclusion
(protection zones, NOTAM-style restrictions, coordination radii), and (3) receiver-side
hardening (filtering, antenna pattern control), \emph{governed by a mandatory
coordination mechanism between the two operating communities rather than a one-time
static exclusion order.}
\end{quote}

The coordination requirement is the pattern's least obvious but most important
component. Neither case was solved by treating the newcomer service as illegitimate:
South Africa's Astronomy Geographic Advantage Act coordinates aviation and telecom
activity within the Karoo rather than banning it, and the 5G C-band resolution
coordinates telecom deployment around airports rather than banning 5G in the band
entirely. Table~\ref{tab:pattern} restates the two case studies against this common
template.

\begin{table*}[!t]
\centering
\caption{The two case studies expressed as instances of the same coexistence design
pattern (Section~\ref{sec:pattern}).}
\label{tab:pattern}
\begin{tabular}{p{3.3cm} p{5.5cm} p{5.5cm}}
\toprule
\textbf{Design-pattern component} & \textbf{Karoo Radio Quiet Zone} & \textbf{5G
C-band / altimeter} \\
\midrule
Frequency separation & Band-specific licensing restrictions within the zone &
$220$\,MHz guard band ($3.98$--$4.2$\,GHz) \\
Spatial/temporal exclusion & Nested coordination radii within
$\sim\!106{,}000\,\mathrm{km^2}$ & Runway-proximate exclusion zones; NOTAM
restrictions \\
Receiver hardening & Site-selection and shielding of the telescope
itself & Fleet-wide altimeter filter retrofit \\
Coordination mechanism & Statutory coordination between SARAO, ICASA, and aviation
authorities & Coordination between FAA/EASA-equivalent regulators and telecom operators \\
Legal instrument & Astronomy Geographic Advantage Act (2007) & Airworthiness
directives; telecom licence conditions \\
\bottomrule
\end{tabular}
\end{table*}

\section{TRANSFERABLE BIG-DATA SIGNAL-PROCESSING TOOLKIT}
\label{sec:toolkit}

The second half of the overlap identified in Section~\ref{sec:intro} concerns
sub-problem (ii): extracting a reliable signal from a large, noisy, RFI-contaminated
data stream in real time. Modern radio interferometers are, first and foremost,
data-engineering systems: MeerKAT's 64 dishes, and the larger SKA array, produce
continuous high-bandwidth voltage streams that must be correlated, calibrated, and
searched for signals that are frequently fainter than the surrounding RFI and
instrumental noise~\cite{Jonas:2016meerkat}. Three techniques from this pipeline have
close counterparts in aviation data problems, though the strength of the analogy varies
technique by technique, as discussed below.

\subsection{RFI FLAGGING AND EXCISION $\leftrightarrow$ ADS-B/RADAR CLUTTER REJECTION}

Automated RFI excision identifies and removes corrupted time-frequency samples before
they contaminate a downstream result. The SumThreshold algorithm, and its
implementation in AOFlagger, formulate this as an iterative thresholding problem on the
time-frequency plane, achieving high recognition accuracy without requiring a
pre-specified data model~\cite{Offringa:2010aof}; more recent work has extended this to
convolutional neural network classifiers operating directly on the same
time-frequency representation~\cite{Akeret:2017mlrfi}. Radar clutter and noise
rejection for weather, terrain, and traffic-separation radar is a close counterpart,
solving essentially the same statistical outlier-flagging problem on a different sensor
modality~\cite{Skolnik:2001radar}. Automatic Dependent Surveillance--Broadcast (ADS-B)
validation is a related but not identical problem: position reports from thousands of
aircraft must be filtered, cross-validated, and screened for spoofed or corrupted
messages in near real time before being trusted for air traffic
control~\cite{Strohmeier:2014adsb}. The screening and cross-validation stage is
directly analogous to RFI excision, but ADS-B security also depends on authentication
and trust mechanisms, since a spoofed message can be statistically indistinguishable
from a genuine one, unlike RFI, which is not adversarially generated. The excision
toolkit therefore transfers to the anomaly-detection layer of ADS-B validation, not to
the full security problem.

\subsection{BEAMFORMING $\leftrightarrow$ PHASED-ARRAY RADAR AND SATCOM}

Beamforming electronically steers an array's combined sensitivity toward a direction of
interest by applying frequency- and geometry-dependent phase weights across array
elements, without any moving parts~\cite{VanTrees:2002array}. The underlying array
mathematics and hardware concept are shared with phased-array weather and surveillance
radar, and with the electronically-steered satellite communication antennas used in
next-generation in-flight connectivity and satcom links. The optimisation objective
differs, however: interferometric beamforming is built around sky localisation and
angular resolution of a static or slowly-moving source field, whereas radar
beamforming is jointly optimised against range-Doppler discrimination and clutter
rejection for fast-moving targets. The transferable asset is the array-processing
formalism itself, not a single unmodified algorithm.

\subsection{STATISTICAL CALIBRATION $\leftrightarrow$ PREDICTIVE MAINTENANCE}

Radio interferometry requires continuous statistical calibration to correct for
instrumental gain drift and environmental effects, so that a weak, real signal is not
absorbed into systematic error. Aircraft predictive-maintenance programmes, which
analyse continuous sensor streams from engines and airframes to detect early signs of
degradation before they manifest as a fault, solve the same statistical problem:
distinguishing a slow, real drift signal from measurement noise in a continuously
streaming, high-volume dataset, an approach formalised in the condition-based
maintenance literature as the joint problem of diagnosing existing faults and
prognosing remaining useful life from continuous multi-sensor data
streams~\cite{Jardine:2006cbm}.

\subsection{MATCHED FILTERING AND ML-ASSISTED DETECTION}

Finally, the detection of a known or partially known signal shape embedded in noise,
whether an astrophysical transient template or a radar return waveform, is classically
solved by matched filtering~\cite{Skolnik:2001radar}, and increasingly augmented by
machine-learning classifiers trained to distinguish genuine signal from interference or
artefact in real time~\cite{Akeret:2017mlrfi}. Table~\ref{tab:toolkit} summarises
the mapping.

\begin{table*}[!t]
\centering
\caption{Shared big-data signal-processing toolkit across radio astronomy and
aviation.}
\label{tab:toolkit}
\begin{tabular}{p{3.4cm} p{5.6cm} p{5.6cm}}
\toprule
\textbf{Technique} & \textbf{Radio-astronomy realisation} & \textbf{Aviation
realisation} \\
\midrule
RFI flagging \& excision & SumThreshold / AOFlagger time-frequency
classification~\cite{Offringa:2010aof} & ADS-B message validation; radar noise/clutter
rejection~\cite{Strohmeier:2014adsb} \\
Beamforming & Electronic steering of interferometer array
sensitivity~\cite{VanTrees:2002array} & Phased-array radar; electronically-steered
satcom antennas \\
Statistical calibration & Continuous gain-drift and environmental
correction & Sensor-drift correction in predictive maintenance \\
ML-assisted detection & CNN-based RFI/signal classification~\cite{Akeret:2017mlrfi}
& Anomaly detection in engine and airframe health monitoring \\
\bottomrule
\end{tabular}
\end{table*}

\section{DISCUSSION}
\label{sec:discussion}

Two implications follow from treating the radio astronomy--aviation overlap as an
engineering problem rather than a coincidence.

\textbf{For regulators}, the coexistence design pattern of Section~\ref{sec:pattern}
generalises beyond the two cases examined here. Any future conflict between a
protected, narrowband, safety- or science-critical receiver and a growing broadband
commercial service, for example uncrewed aerial vehicle (UAV) telemetry and control
links operating near radio-quiet zones, or next-generation 6G deployments approaching
other protected bands, where large-scale terrestrial sub-THz networks have already been
shown to introduce harmful RFI into passive Earth-sensing and radio astronomy
systems~\cite{Testolina:2024passive}, can be approached with the same three-lever,
coordination-first template that both the Astronomy Geographic Advantage Act and the 5G
C-band resolution independently arrived at. Legislating this pattern proactively, rather
than negotiating it reactively after a conflict emerges, is the clearest lesson
transferable from the Karoo case to future spectrum disputes. A fourth lever is
beginning to emerge alongside the three identified in Section~\ref{sec:pattern}:
real-time, data-driven coordination in place of static exclusion, as demonstrated by
recent telescope boresight-avoidance trials between the Green Bank Observatory and the
Starlink satellite constellation, in which the telescope's live pointing and observing
frequency are shared with the satellite operator so that individual downlink beams can
be redirected or disabled during a close pass~\cite{Nhan:2024starlink}. This points
toward coexistence frameworks that adjust $M(f,d)$ dynamically, rather than fixing it by
a single legislated boundary.

\textbf{For infrastructure and workforce planning}, the toolkit mapping in
Section~\ref{sec:toolkit} implies that RFI-mitigation software and hardware developed
for radio astronomy is a directly reusable asset for aviation, telecommunications, and
defence radar sectors, and that a national investment in radio-astronomy data-science
training (motivated by projects such as MeerKAT and the SKA) simultaneously builds a
workforce pipeline serving these adjacent sectors, rather than requiring each to build
a duplicate, siloed training track.

Two limitations of the present analysis are worth stating plainly, rather than leaving
them implicit in the caveats scattered through earlier sections.

\textbf{Simplified link-budget model.} The coexistence-margin formalism of
Section~\ref{sec:framework} is a line-of-sight link budget; real coordination
decisions in both domains rely on detailed propagation modelling, statistical
exceedance criteria, and case-specific antenna patterns that sit outside
equations~\eqref{eq:margin}--\eqref{eq:fspl}. This is why we report $M(f,d)$ as a
structural quantity rather than as a fitted number for either case study: the
framework identifies which levers matter and how they trade off, not the precise
coordination distance a regulator would set.

\textbf{Illustrative, not quantitative, schematics.} The figures in
Sections~\ref{sec:framework} and~\ref{sec:karoo} are illustrative representations of
documented regulatory and physical structures, not quantitative reproductions of any
specific coordination study. We have attributed every numerical threshold quoted in
the text to its originating standard or report rather than deriving new ones, so the
schematics should be read as structural diagrams of the pattern, not as calibrated
predictions.

\section{CONCLUSION}
\label{sec:conclusion}

Radio astronomy and aviation solve versions of the same two engineering problems: how
to protect a scarce, invisible resource from interference, and how to extract a
reliable signal from an overwhelming, noisy data stream fast enough for the result to
matter. This paper has made that overlap precise rather than anecdotal, in three ways:

\begin{enumerate}[label=\textbf{(\roman*)}, leftmargin=*, itemsep=6pt]
  \item \textbf{A unified coexistence-margin framework} (Section~\ref{sec:framework})
  that expresses the ITU-R RA.769 radio astronomy protection criterion and the RTCA
  radar-altimeter interference threshold as the same underlying quantity,
  $M(f,d)=\Pth(f)-\Prx(f,d)$, evaluated with domain-specific definitions of $\Pth$.
  \item \textbf{Two case studies} (Sections~\ref{sec:karoo}--\ref{sec:cband}) showing
  that the Karoo Radio Quiet Zone and the 5G C-band/altimeter resolution, despite
  independent regulatory origins, converged on the same three-lever coexistence
  strategy frequency separation, spatial/temporal exclusion, and receiver
  hardening governed by mandatory coordination rather than static exclusion,
  distilled into a general \textbf{coexistence design pattern} (Section~\ref{sec:pattern}).
  \item \textbf{A transferable signal-processing toolkit} (Section~\ref{sec:toolkit})
  relating RFI excision, beamforming, statistical calibration, and matched
  filtering/ML-assisted detection, as developed for radio interferometry, to their
  closest counterparts in ADS-B validation, phased-array radar, and predictive
  maintenance, distinguishing the techniques that transfer directly (statistical
  calibration, matched filtering, radar clutter rejection) from those that transfer
  only in part, alongside a domain-specific component (the shared array-processing
  formalism behind beamforming; the anomaly-detection layer, but not the
  authentication layer, of ADS-B validation).
\end{enumerate}

The practical conclusion is that neither field needs to solve these problems from
first principles in isolation. A regulatory template, an engineering design pattern,
and a software/hardware toolkit already exist on one side of this overlap and are
directly reusable on the other; the opportunity is active collaboration, shared
monitoring infrastructure, a shared regulatory template, and a shared technical
workforce, rather than parallel, siloed development in two communities solving the
same problem twice.

A few directions follow naturally from here: (i) a quantitative, propagation-model-based
evaluation of $M(f,d)$ for a specific coordination scenario, rather than the
structural treatment given here; (ii) extension of the design pattern to further
narrowband-versus-broadband conflicts, such as UAV telemetry links or passive
Earth-sensing bands approached by 6G deployment; (iii) a systematic comparison of the
dynamic, data-driven coordination lever demonstrated by GBT--Starlink boresight
avoidance against the static levers used in both case studies examined here; and (iv)
an empirical audit of which specific RFI-mitigation software components (e.g.\
AOFlagger-class flaggers) are already in use, or could be adopted with minimal
modification, across ADS-B validation and phased-array radar pipelines.

\newpage
\section*{DATA AVAILABILITY}
This paper synthesizes publicly available regulatory, standards, and literature sources, all of which are cited in the references below; no new datasets were generated.

\begin{IEEEbiography}[{\includegraphics[width=1in,height=1.25in,clip,keepaspectratio]{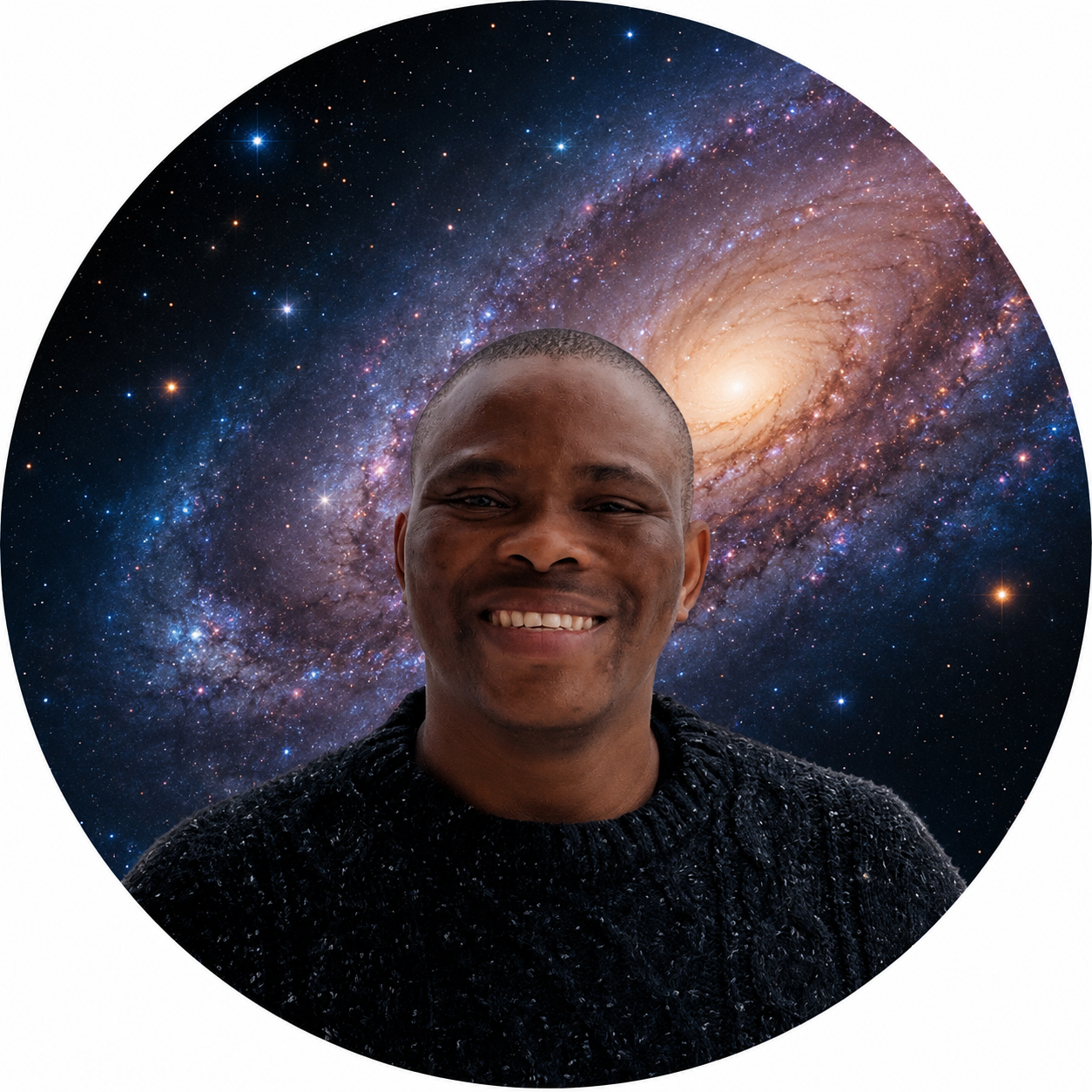}}]{AUTHOR }\; \noindent\textbf{Dr. Dlamini} received the Ph.D. degree in astrophysics from the University of the Western Cape, Cape Town, South Africa, in 2024, the M.Sc. degree in applied mathematics from the University of Cape Town, Cape Town, South Africa, and the B.Sc.(Hons.) degree in astrophysics, also from the University of Cape Town. He is currently a Senior Lecturer of Astronomy with the University of South Africa, Johannesburg, South Africa. He is a radio astronomer and cosmologist whose work uses HI (neutral hydrogen) 21\,cm intensity mapping and large-scale galaxy surveys to probe cosmic structure formation and the physics of the early Universe. He was previously a SARChI Research Fellow in Astronomy with the University of Cape Town, where he conducted research on large-scale structure analysis using MeerKAT and PHANGS Galaxies data. He is the author or co-author of more than a dozen peer-reviewed journal articles, including the co-authored paper ``Constraining the Growth Rate on Linear Scales by Combining SKAO and DESI Surveys'' (\textit{Eur. Phys. J. C}, 2024). His solely authored work, ``Fast Radio Burst Dispersion Measure--Timing Cross-Correlations: Bias Self-Calibration and Primordial Non-Gaussianity Constraints'' (\textit{Eur. Phys. J. C}, 2026), presents a self-calibrating framework for extracting primordial non-Gaussianity signatures from FRB dispersion measure data. His research interests include neutral hydrogen intensity mapping, radio astronomy, radio cosmology, large-scale cosmic structure, and galaxy clustering. He also maintains a strong personal interest in aviation. Dr.\ Dlamini is a Junior Member of the International Astronomical Union (IAU), a member of the PHANGS Collaboration, and a member of the African Astronomical Society (AfAS). He was a Finalist for the Research Software Award at the 2025/2026 NSTF-South32 Awards and served as a Session Convenor at the SCAR Open Science Conference in Oslo, Norway, in 2026.
\end{IEEEbiography}

\end{document}